\documentclass[usenatbib]{mn2e}
\usepackage{natbib}
\usepackage{amsmath}
\usepackage{epsfig}

\def\reff@jnl#1{{\rm#1\/}}

\def\aj{\reff@jnl{AJ}}                  
\def\araa{\reff@jnl{ARA\&A}}            
\def\apj{\reff@jnl{ApJ}}                        
\def\apjl{\reff@jnl{ApJ}}               
\def\apjs{\reff@jnl{ApJS}}              
\def\ao{\reff@jnl{Appl.Optics}}         
\def\apss{\reff@jnl{Ap\&SS}}            
\def\aap{\reff@jnl{A\&A}}               
\def\aapr{\reff@jnl{A\&A~Rev.}}         
\def\aaps{\reff@jnl{A\&AS}}             
\def\azh{\reff@jnl{AZh}}                        
\def\baas{\reff@jnl{BAAS}}              
\def\jrasc{\reff@jnl{JRASC}}            
\def\memras{\reff@jnl{MmRAS}}           
\def\mnras{\reff@jnl{MNRAS}}            
\def\physrep{\reff@jnl{Phys.Rep.}}
\def\pra{\reff@jnl{Phys.Rev.A}}         
\def\prb{\reff@jnl{Phys.Rev.B}}         
\def\prc{\reff@jnl{Phys.Rev.C}}         
\def\prd{\reff@jnl{Phys.Rev.D}}         
\def\prl{\reff@jnl{Phys.Rev.Lett}}      
\def\pasp{\reff@jnl{PASP}}              
\def\pasj{\reff@jnl{PASJ}}              
\def\qjras{\reff@jnl{QJRAS}}            
\def\skytel{\reff@jnl{S\&T}}            
\def\solphys{\reff@jnl{Solar~Phys.}}    
\def\sovast{\reff@jnl{Soviet~Ast.}}     
\def\ssr{\reff@jnl{Space~Sci.Rev.}}     
\def\zap{\reff@jnl{ZAp}}                        
\def\nat{\reff@jnl{Nature}}             

\def\sun{\hbox{$\odot$}}

\newcommand{\hkpc}{h^{-1}\mathrm{kpc}}
\newcommand{\hMsun}{h^{-1}M_{\odot}}

\newcommand{\Omegam}{\Omega_{m}}

\newcommand{\be}{\begin{equation}}
\newcommand{\ee}{\end{equation}}
\newcommand{\bea}{\begin{eqnarray}}
\newcommand{\eea}{\end{eqnarray}}

\title[Galaxy-galaxy lensing]{Galaxy-galaxy lensing:
  dissipationless simulations versus the halo model} 

\author[Mandelbaum et al.]{
Rachel Mandelbaum$^1$\thanks{Electronic address:
    {\tt rmandelb@princeton.edu}},
Argyro Tasitsiomi$^2$,
Uro\v s Seljak$^{1,3}$, Andrey V. Kravtsov$^{2,4}$,
\newauthor \hspace{0.009in}  Risa H. Wechsler$^{2,5}$
\\$^1$Department of Physics, Princeton University, Princeton, NJ 08544,
USA 
\\$^2$Department of Astronomy \& Astrophysics, Kavli Institute for Cosmological Physics, The University of Chicago, Chicago, IL 60637, USA
\\$^3$International Centre for Theoretical Physics, Strada Costiera
11, 34014 Trieste, Italy
\\$^4$Enrico Fermi Institute, The University of Chicago 
\\$^5$Hubble Fellow, Enrico Fermi Fellow}
\date{\today}

\begin{document}
\maketitle

\begin{abstract}
Galaxy-galaxy lensing is a powerful probe of the relation between 
galaxies and dark matter halos, but its theoretical interpretation requires 
a careful modeling of various contributions, such as the contribution 
from central and satellite galaxies. 
For this purpose, a phenomenological approach based on the halo model 
has been developed, allowing for fast exploration of the parameter space
of models. In this paper, we investigate
the ability of the halo model to extract information from the g-g weak
lensing signal by comparing 
it to high-resolution dissipationless simulations that resolve subhalos. 
We find that the halo model reliably determines 
parameters such as the host halo mass of central galaxies, the fraction of 
galaxies that are satellites, and their radial distribution inside
larger halos. If there is a significant scatter present in the central 
galaxy host halo mass distribution, then
the mean and median mass of that distribution can differ significantly 
from one another, 
and the halo model mass determination lies between the two. 
This result suggests that when analyzing the data, 
galaxy subsamples with a narrow central galaxy
halo mass distribution, such as those based on stellar mass,
should be chosen for a simpler interpretation of the results. 

\end{abstract}


\setcounter{footnote}{0}

\section{Introduction}

Understanding the connection between the spatial distribution of galaxies
and dark matter (DM) is one of the most important problems in modern
cosmology.  From the perspective of fundamental physics, this
connection needs to be understood if we wish to use galaxy clustering
to determine dark matter correlations, a potentially powerful
discriminator between cosmological models. This connection can also be
a powerful test of the tenets of the standard cosmological paradigm,
such as the collisionless nature of cold dark matter.  From the
astrophysics perspective, this connection is an essential ingredient
in the physics of galaxy formation. Current models based on pure
N-body simulations, semi-analytic models or hydrodynamic simulations
can accomodate many observational aspects of this connection, but
several ingredients of these models remain uncertain and need to be
inserted ad-hoc, so it is not clear how much of the success is a
result of the allowed freedom within these models. It is important to
look for new ways to test this connection, to confirm and improve the
existing models of structure formation, and to distinguish
astrophysical effects from cosmological ones.

One of the probes of the galaxy-DM connection that recently became 
available is weak lensing around 
galaxies, or galaxy-galaxy (hereinafter g-g) lensing 
 \citep{1984ApJ...281L..59T,1996ApJ...466..623B,1998ApJ...503..531H,2000AJ....120.1198F,2001ApJ...551..643S,2001astro.ph..8013M,2003MNRAS.340..609H,2004AJ....127.2544S,2004ApJ...606...67H,2005PhRvD..71d3511S}. 
Gravitational lensing induces
tangential shear distortions of background galaxies 
around foreground galaxies, allowing
direct measurement of the galaxy-dark matter correlation function around 
galaxies. 
The individual distortions are small (of order 0.1\%), but by 
averaging over all foreground galaxies within a given subsample, 
we obtain high signal 
to noise in the shear as a function of angular separation from the galaxy. 
If we know the galaxy redshifts,  
the shear signal
can be related to the projected mass 
density as a function of proper distance from the galaxy. 
This allows us to determine statistically the dark matter distribution 
around any given galaxy sample. 

In recent years, the progress on the observational 
side of g-g lensing has been remarkable. 
In the latest Sloan Digital Sky Survey (SDSS) 
analyses \citep{2004AJ....127.2544S,2005PhRvD..71d3511S}, 
20-30 sigma detections of the signal as a function of physical 
separation have been obtained. Similarly high S/N detections 
have also been observed as a function of angular separation 
with other surveys \citep{2004ApJ...606...67H}. This increased statistical 
power has been accompanied by a more careful investigation of systematic
errors, such as calibration biases and intrinsic correlations, 
which for the SDSS are currently around 10\% and therefore 
already dominate the error budget \citep{2004MNRAS.353..529H}. 
As the quality and 
quantity of the data and its analysis improves, the reliability of 
theoretical interpretation must be improved as well. The
goal of the present paper is to compare the various 
theoretical analyses among themselves, and to discuss how they
should be applied to the data.

Theoretical analysis of g-g lensing falls into two categories. The first approach 
is a direct comparison of simulations 
to the data \citep{2001MNRAS.321..439G,2003MNRAS.339..387Y,2004ApJ...601....1W,2004ApJ...614..533T}. This approach 
is direct, but rather expensive, since the process 
of galaxy formation is not understood sufficiently well to result in 
unique predictions for a given cosmological model, nor is the cosmological 
model  itself
determined yet. 
Moreover, current simulations still suffer from a limited 
dynamical range, in the sense that they require a high mass and force 
resolution to resolve individual galaxies and their associated dark matter 
halos, while at the same time they must also have 
sufficiently large volume to simulate 
a representative region of the universe. 
Several simulations of varying box size are needed 
to cover the whole observational range in luminosity and scale. 
As a result, many different simulations would be needed to explore the
whole range of parameter space; at present, only a handful of 
simulations have been used for this application. 

The second approach is to use the halo model \citep[for a review, see][]{2002PhR...372....1C}
to describe the connection between 
galaxies and dark matter \citep[][hereinafter GS02]{2002MNRAS.335..311G}. This approach is more phenomenological, 
and for a given galaxy class 
leads to determination of quantities such as the virial mass distribution and
the fraction of these galaxies that are 
satellites, which are useful quantities for constraining 
the galaxy formation models and cosmological models. 
While the information extracted from the data using the halo model may be all 
that is needed to quantify the galaxy-dark matter connection, 
halo models are reliable only to the extent that they are able to 
reproduce the simulations, so they must be tested and calibrated on 
the simulations before applications to the real data are reliable.

In this paper, we attempt
to understand how well the phenomenological halo model developed in~\cite{2000MNRAS.318..203S} and studied in GS02
 can
reproduce the simulations. We are interested in its ability to 
reproduce the g-g lensing signal, which in turn can be used to 
determine the halo mass probability distribution, 
the radial profile of satellite galaxies inside larger halos, and
other quantities of interest. 

\section{Simulations}
\label{sec:sims}

We use simulations described in 
\cite{2004ApJ...614..533T}, where the reader can find further details.
We assume the concordance flat
$\Lambda$CDM model: $\Omegam=1-\Omega_{\Lambda}=0.3$, $h=0.7$, where
$\Omegam$ and $\Omega_{\Lambda}$ are the present-day matter and
vacuum densities, and $h$ is the dimensionless Hubble constant defined
as $H_0\equiv 100h{\ }{\rm km\ s^{-1}\,Mpc^{-1}}$. The power spectrum
normalization is $\sigma_8=0.9$. The model is 
consistent with recent observational constraints
\citep[e.g.,][]{2003ApJS..148..175S,2004PhRvD..69j3501T}. The effects
of the power spectrum normalization, box size and
cosmic variance were studied in \cite{2004ApJ...614..533T} using a range of simulation
box sizes. Given current constraints, 
it is impossible to achieve the desired level of precision 
across the entire range of luminosities probed by the observations. 
Small box sizes can achieve sufficient mass resolution to resolve very 
small halos, but the sampling variance due to large scale fluctuations 
is large, which results in large errors on predictions at scales above 
a few hundred kpc. To reduce sampling errors, we will choose the largest box 
available from  the simulations in \cite{2004ApJ...614..533T}, 
a 120 $h^{-1}$Mpc box with $512^3$ particles. The particle mass of this 
simulation is $m_p=1.07\times 10^9\hMsun$, so only halos with virial mass 
above a few times $10^{11}h^{-1}M_{\sun}$ are resolved. For this reason, we 
will restrict the analysis to galaxies brighter than $M_r=-19$.

The simulations were run using the Adaptive Refinement Tree $N$-body
code \citep[ART;][]{1997ApJS..111...73K,1999ApJ...520..437K}. The ART code reaches
high force resolution by refining all high-density regions with an
automated refinement algorithm. The criterion for refinement is the
mass of particles per cell. 
The initial grid is $1024^3$ and the
refinement criterion is level- and time-dependent. At the early stages
of evolution ($a<0.65$) the thresholds are set to 2, 3, and 4 particle
masses for the zeroth, first, and second and higher levels,
respectively. At low redshifts, $a>0.65$, the thresholds for these
refinement levels are set to 6, 5, and 5 particle masses.  The lower
thresholds at high redshifts are set to ensure that collapse of
small-mass halos is followed with higher resolution. The maximum
achieved level of refinement is $L_{\rm max}=8$. As a function of redshift the
maximum level of refinement is equal to $L_{\rm max}=6$ for $5<z<7$,
$L_{\rm max}= 7$ for $1<z<5$, $L_{\rm max}\geq 8$ for $z<1$. The peak
formal resolution is $h_{\rm peak}\leq 1.8\hkpc$ (physical). 

A variant of the Bound Density Maxima halo finding algorithm
\citep{1999ApJ...516..530K} is used to identify halos and the subhalos
within them.  The details of the algorithm and parameters used in the
halo finder can be found in \citet{2004ApJ...609...35K}. The main
steps of the algorithm are the identification of local density peaks
(potential halo centers) and analysis of the density distribution and
velocities of the surrounding particles to test whether a given peak
corresponds to a gravitationally bound clump. More specifically,
density, circular velocity, and velocity dispersion profiles are
constructed around each potential halo center. The
unbound particles are then removed iteratively using the procedure outlined in
\citet{1999ApJ...516..530K}.  The final profiles are constructed using
only bound particles. We use these profiles to calculate properties of
halos, such as the circular velocity profile $V_{\rm
  circ}(r)=\sqrt{GM(<r)/r}$, and compute the maximum circular velocity
$V_{\rm max}$.  

In this study, we distinguish between {\it host halos} with centers
that do not lie within any larger virialized system, and {\it subhalos}
(or {\it satellites}) located within the virial radii of larger systems. 
We associate the former with central galaxies and the latter with 
non-central galaxies or satellites. 
To classify the halos, we calculate the formal
boundary of each object as the radius corresponding to an enclosed
overdensity of 180 with respect to the mean density of the universe.
Note that we do not consider the center of a host halo to 
be a subhalo. Thus, host halos may or may not contain any subhalos
with circular velocity above the threshold of a given sample. The host
centers, however, are included in clustering statistics because we
assume that each host harbors a {\it central} galaxy at its center.
Therefore, the total sample of galactic halos contains central and
satellite galaxies. The former have the positions and maximum circular
velocities of their host halos, while the latter have the positions and
maximum circular velocities  of subhalos.
In a cluster, for
example, the brightest central galaxy that typically resides near the
center would be associated with the cluster host halo in our
terminology. All other galaxies within the virial radius of the
cluster would be considered ``satellites'' associated with subhalos.

Our galaxy sample is created by assigning realistic
SDSS luminosities and colors to dark matter halos.  To construct mock
galaxy catalogs for comparison with observations, one must define
selection criteria for particular halo properties to mimic the
selection function of the observational sample as closely as possible.
Halo mass is often used to define halo catalogs; e.g., a catalog can
be constructed by selecting all halos in a given mass range.  However,
the mass and radius are poorly defined for the satellite halos due to
tidal stripping, which alters a halo's mass and physical extent
\citep[see][]{1999ApJ...516..530K}.  Therefore, we use the maximum circular
velocity, $V_{\rm max}$, as a proxy for the halo mass.  For isolated
halos, $V_{\rm max}$ and the halo's virial mass are directly related.
For subhalos, $V_{\rm max}$ will experience secular decrease but at a
relatively slow rate \citep{2004ApJ...609..482K}.

To mimic the observational selection function, $r$-band luminosities
are assigned to the halos as follows. We match the cumulative velocity
function $n(>V_{\rm max})$ of the halos to the SDSS observed $r$-band
cumulative luminosity function \citep{2003ApJ...592..819B}.  Note that
$n(>V_{\rm max})$ includes both isolated host halos and subhalos.  We
use the $r$-band data since that band was used for SDSS spectra
selection, has a more reliable luminosity function
measurement observationally, and is the focus of most SDSS analyses. We
first obtain the average $V_{\rm max} -M_{r}$ relation by matching
$n(>V_{\rm max})$ to $n(<M_{\rm r})$.  This was the same method used
to assign luminosities to subhalos in \citet{2004ApJ...609...35K}, in
which galaxy clustering properties were reproduced remarkably well.
The mean redshift of the lens galaxies is $0.1$, so
we use halo catalogs at this redshift.  

One may also introduce scatter in the relation between $V_{\rm max}$
and $M_r$, which here we have chosen as a Gaussian in $M_r$ at fixed $V_{\rm max}$,
as described in detail in \cite{2004ApJ...614..533T}.  The amplitude
and dependence of the scatter on galaxy luminosity or halo mass are
very uncertain, so here we will simply explore this example to
investigate its consequences. The value of scatter is meant to be
realistic for the current data, which are a mixture of early and late
type galaxies and for which intrinsic reddening corrections have not
been applied.  The Gaussian scatter is introduced in a fashion that
keeps the luminosity function constant, which results in somewhat
lower scatter for higher luminosity galaxies and vice versa.  Below we
present extensive comparisons of the halo model and the actual
simulation results with and without scatter.

\section{Overview of the halo model \label{theory}}

\subsection{Dark matter halos}

In current cosmological models, structure grows hierarchically 
from small, initially Gaussian fluctuations. Once the 
fluctuations go nonlinear, they collapse into virialized halos. 
The spatial density of halos 
as a function of their mass $M$ is specified
by the halo mass function $dn / dM$, which in general is a function of 
redshift $z$. It can be written as
\begin{equation}
\frac{dn}{dM} dM=\frac{\bar{\rho}}{M}f(\nu)d\nu,
\end{equation}
where $\bar{\rho}$ is the mean matter density of the universe. We
introduced the
function $f(\nu)$, which can be
expressed in units in which it has a theoretically universal 
form independent of the power spectrum or redshift if written
as a function of peak height
\begin{equation}
\nu=\left[\frac{\delta_c}{\sigma(M)}\right]^2.
\label{nu}
\end{equation}
Here $\delta_c=1.686$ is the linear overdensity at which a spherical perturbation
collapses at redshift $z$, and
$\sigma(M)$ is the rms fluctuation in spheres that contain on average
mass $M$ at an initial time, extrapolated using linear theory to $z$.

The first analytic model for the mass function was 
proposed by \cite{1974ApJ...187..425P}. 
While it correctly predicts the abundance of
massive halos, it overpredicts the abundance of halos around 
and below the nonlinear mass scale $M_{\rm nl}(z)$ (defined below). 
An improved version has been proposed by \cite{1999MNRAS.308..119S},
\begin{equation}
\nu f(\nu)=A(1+ \nu'^{-p})\nu'^{1/2} e^{-\nu'/2},
\label{fnu}
\end{equation}
where $\nu'=a\nu$, with $a=0.707$ and $p=0.3$ as the best fit values
(the \cite{1974ApJ...187..425P} mass function corresponds to $a=1$, $p=0$).
Further modifications to this expression 
have been suggested in \cite{2004ApJ...605..709Y}, but the
effects are very small and we will ignore them here. 
The constant $A$ is determined by mass conservation,
requiring that the integral over the mass
function times the mass gives the mean density or, equivalently, 
$\int f(\nu)d\nu=1$.  
It has been shown that 
the form in equation \ref{fnu} 
can be 
derived analytically within the framework of the ellipsoidal collapse model
\citep{2001MNRAS.323....1S}.   For a given peak height, we can also compute the bias 
\begin{equation}\label{bias}
b(\nu) = 1+\frac{\nu'-1}{\delta_c}+\frac{2p}{\delta_c(1+\nu'^p)},
\end{equation}
where for the purpose of computing the bias we use $a=0.73$ and
$p=0.15$ rather than the values from the mass function, in order to
best match the results from~\cite{2004MNRAS.355..129S}.  As shown in
Fig.~2 of \cite{2005astro.ph..6395W}, the original
\cite{1999MNRAS.308..119S} mass function and this modified one give
nearly identical results in the most relevant range of halo masses,
$10^{11}$--$10^{14}$ $h^{-1}M_{\odot}$.  In support of this assertion,
we note that when we compare the predicted halo model lensing signal
computed with the two mass functions, the difference is less than 1
per cent.

The halo mass is defined in terms of 
the linking length parameter of the friends-of-friends (FoF) algorithm, 
which is 0.2 for the simulations used in \cite{1999MNRAS.308..119S}. 
This definition roughly corresponds to spherical overdensity halos of 180 times the 
background density \citep{2001MNRAS.321..372J,2001A&A...367...27W}. For the range of
masses of interest here and their corresponding halo concentrations,
it is about 30-35\% larger
than the mass defined as the mass within the radius where the 
density is 200 times the 
critical density. 
Since we use $\Omega_m=0.3$ when computing the 
virial masses, they are defined as the mass within the 
radius in which the mean 
density is 54 times the critical density. 
We will also define the concentration parameter relative to this 
radius. 

\subsection{Halo-galaxy connection and galaxy-galaxy lensing}\label{SS:hgconn}

In the current paradigm of structure formation, all galaxies form inside 
dark matter halos. 
While this is generally accepted, 
we also know that the relation between the two is not one-to-one, and
galaxies of the same 
luminosity can be found in halos of different masses. For example, a 
typical galaxy like the Milky Way 
may be found at the center of a low 
mass halo with a typical size 200 kpc, it may be part of a small 
group with typical size 500 kpc or it 
may be a satellite in a cluster with a typical size of 
1-2 Mpc.  
We wish therefore to determine the probability $\Delta P$
for a galaxy in this sample to be in a halo of mass $M \pm \Delta M/2$. 
To describe this probability, we will use the conditional halo mass  probability
distribution $dP/dM \equiv p(M;{\rm L})$ (GS02).
With g-g lensing, one can in principle determine the full halo mass function, 
since small halos contribute only at small scales, while large halos such 
as clusters also generate signal 
at larger scales.
The fact that g-g lensing measures the signal over a wide range of
scales facilitates the 
determination of the full halo mass function for a given subsample.
In practice, given the measurement errors, a 
model must be adopted to extract this information from the data. 

Galaxy-galaxy lensing measures the tangential shear 
distortions in the shapes of background galaxies 
induced by the mass distribution around foreground galaxies.
Because the shear distortions $\gamma_t$ are very small, in our case $10^{-3}$, 
while the typical galaxy shape noise is 0.3, we
must average over many foreground-background pairs to extract the signal. 
The result is a measurement of the shear-galaxy cross-correlation as a 
function of relative foreground-background separation on the sky. 
We will assume that the redshift of the foreground galaxy 
is known, so  one can express the relative 
separation in terms of transverse comoving scale $R$. 
If, in addition, the
redshift distribution of the background galaxies, or their actual 
redshifts, are known, then one can relate the shear distortion $\gamma_t$ to
$\Delta\Sigma(R)=\bar{\Sigma}(<R)-\Sigma(R)$, 
where $\Sigma(R)$ is the 
surface mass density at the transverse separation $R$ and $\bar{\Sigma}(<R)$ 
its mean within $R$, via
\be
\gamma_t=\frac{\Delta\Sigma(R)}{\Sigma_{\rm crit}}.
\ee
Here 
\be
\Sigma_{\rm crit}=\frac{c^2}{4\pi G} \frac{r_S}{(1+z_L)r_L r_{LS}},
\label{Sigma}
\end{equation}
where $r_L$ and $r_S$ are the comoving distances to the
lens and source, respectively, $r_{LS}$ is the comoving
distance between the two and $z_L$ is the redshift 
of the lens. If only the probability distribution for 
source redshifts is known, then this expression needs to be integrated 
over. In principle, the relation between 
comoving distance and measured redshift depends on cosmology, 
but since we are dealing with low redshift objects,
varying cosmology within the allowed range makes little difference.
We will assume a cosmology with $\Omega_m=0.3$ and 
$\Omega_{\Lambda}=0.7$, and we work in comoving coordinates throughout
the paper. 
 
We will first overview the halo model formalism of GS02, 
beginning the discussion with a simplified description. 
Let us assume that a given halo of mass $M$ produces 
an average lensing profile $\Delta\Sigma(R,M)$. This can be obtained via
line of sight integration over the
dark matter profile, which in this paper is modeled as a NFW profile 
\citep{1996ApJ...462..563N} 
\begin{equation}
\rho(r)=\frac{\rho_s}{(r/r_s)(1+r/r_s)^{2}}.
\label{rho}
\end{equation}
This model assumes that the profile shape is
universal in units of the scale radius $r_s$, while its characteristic density
$\rho_s$  or concentration $c_{\rm dm}=r_v/r_s$ may depend on the halo mass,
which here will be modeled as 
\citep{2001MNRAS.321..559B,2001ApJ...554..114E} 
\be
c_{\rm dm}= 10\left(\frac{M}{M_{\rm nl}(z)}\right)^{-0.13}. 
\label{cm}
\ee
$M_{\rm nl}(z)$, which depends on the cosmology, is defined such
that the rms linear density fluctuation, extrapolated to redshift $z$,
within a sphere containing mass $M_{\rm nl}$ is equal to $\delta_c$.
Since most of the signal is at $R>50-100$ $h^{-1}{\rm kpc}$, baryonic 
effects can be neglected \citep[see, e.g.,][]{2004ApJ...616...16G}, 
dark matter profiles are well 
determined from simulations, and concentration or the choice of the 
halo profile does not play a major role. 
The average g-g lensing signal for 
a galaxy with luminosity $L$ is 
\begin{equation}
\langle \Delta\Sigma \rangle (R;L)=\int p(M;L)\Delta\Sigma(R,M)dM. 
\label{dsp}
\end{equation}
Because of noise in the g-g lensing signal, we cannot invert the 
relation to obtain the
conditional mass probability 
distribution with arbitrary precision, but instead assume some 
functional form for it and fit for its parameters. 

Our simplified description so far ignores the fact that there are 
two distinct galaxy types that need to be modeled separately. The first
type are the galaxies that formed at the centers of dark matter halos, such 
as the so-called field 
galaxies or cD galaxies in the cluster centers. The second type are the non-central 
galaxies, or satellites.
We know that a galaxy of a given 
luminosity can be of either type, so following GS02,
we split $p(M;L)$ into two parts, $p^{\rm C}$ and $p^{\rm NC}$, 
representing respectively central and non-central galaxies, 
with the fraction of non-central galaxies in each luminosity 
bin $L_i$ given by a free parameter $\alpha_i$:
\begin{equation}
p(M;L_i) = (1-\alpha_i)~p^{\rm C}(M;L_i)+\alpha_i~p^{\rm NC}(M;L_i)~.
\end{equation}
$\alpha_i$ is, by definition, the satellite fraction determined for
the given lens sample as a whole.

For the central galaxy population 
we assume that 
the relation between the halo mass 
and galaxy luminosity is tight and we model this 
component with a delta-function,
\begin{equation}
p^{\rm C}(M;L_i)dM = \delta^D(M-M_{0,i})dM~,
\end{equation}
where $M_{0,i}$ is the typical halo mass of the $i$th luminosity bin. 
In reality, this component should have some width both because 
of intrinsic scatter in the $M-L$ relation, scatter due to inclination and dust,
 and because 
we work with luminosity bins of finite width. 
Explicit tests have shown that the results are only weakly affected
even if the scatter is a factor of 2-3 
in mass (GS02). If the scatter is larger, then 
the effects can be larger;   
we will use simulations to investigate this effect. 

The non-central galaxies are different in that they have presumably formed 
in smaller halos which then merged into larger ones. It is thus reasonable
to assume that their luminosity
is not related to the final (host) halo mass. Instead we 
assume a relation between the number of these non-central galaxies
and the halo mass:
the larger the halo, the more satellites of a given luminosity 
one expects to find in it. We assume
this relation is a power law, $\langle N \rangle (M;L) \propto M^{\epsilon}$,
above some minimal halo mass $M_{\rm min}$, 
which should be larger than the halo mass of the central galaxy 
component above, since we are assuming that there is already another
galaxy at the halo
center.  Below this cutoff, the number of 
galaxies quickly goes to zero. 
These assumptions imply
\begin{equation}
p^{\rm NC}(M;L_i) dM \propto F(M) 
M^{\epsilon_i} \frac{dn}{dM}dM~.
\end{equation}
GS02 used $F(M)= \Theta^H(M-M_{{\rm min},i})$ where $M_{{\rm min},i}=3
M_{0,i}$; here, we will use a slightly more realistic functional form
that is smoother than the step function and matches the simulations
better. This is described in more detail in the next section.  We have
verified that the two expressions do not yield significantly different 
results.  Semi-analytic models of galaxy formation
\citep[][GS02]{1999MNRAS.303..188K}, subhalos in N-body simulations
\citep{2004ApJ...609...35K}, and direct observational measurements
\citep{2004ApJ...610..745L,2004astro.ph..8569Z} agree with this model, and predict that for most galaxies,
$\epsilon \sim 1$ and $\alpha \sim 0.2$. In fact, when attempting to fit for both
$\epsilon$ and $\alpha$, we find that these two parameters  are
almost completely degenerate. We therefore assume fixed  $\epsilon=1$, and do not fit for
it throughout this study.  

For the non-central component, the weak lensing profile
$\Delta\Sigma(R,M)$ is a convolution of the halo profile with the
radial distribution of the galaxies.  Since we are explicitly
excluding the central galaxies, the non-central galaxy component of
the g-g lensing signal does not peak at the center. Instead, for a
given halo mass, it is small at small radii, peaks at a fraction of
the virial radius, and drops off at large radii.

We assume that the radial distribution of galaxies is proportional to
the dark matter profile, $c_g=ac_{\rm dm}$.  Observationally, there is
not much evidence for any departures from $a=1$, but the constraints
are weak and only exist for clusters \citep{1997ApJ...478..462C}.  As
a result, $a$ could differ significantly from unity, as discussed in
more detail in \cite{2005ApJ...618..557N}.  We will show that g-g
lensing may be a useful way to measure it observationally for groups
as well.  

On scales below $2h^{-1}$Mpc, the one-halo or Poisson term dominates
the g-g lensing signal.  As shown in GS02, the lensing signal from
neighboring halos (called the ``halo-halo'' or hh term) can be
neglected for $R<2h^{-1}$Mpc, while on very large scales this term
dominates and follows linear theory.  For the sake of consistency
between results on large and small scales, we have included it here.
Since the NFW profile is not abruptly truncated at the virial radius
in simulations, we have investigated the dependence of the results on
the maximum radius to which the halo profile is integrated in, e.g.,
equation~\ref{ykm}. We define this radius in units of the virial
radius, so $x_1=1$ means that no mass beyond the virial radius is
taken into account. (This is only true for the non-central term; when
computing the Poisson term due to the central halo, the profile is not
truncated at $x_1 r_{vir}$ when projecting to get $\Sigma(R)$ from $\rho(r)$.)  Since we do not assume that the halo
mass function accounts for all the mass in the universe, there is no
concern of double-counting mass when using $x_1 > 1$.  The halo-halo correlations also depend on the
large scale bias, which for a given halo mass distribution is given by
equation~\ref{bias}.  Unfortunately, the small box size of the
simulations implies that the sampling variance is important on large
scales, so the errors on large scales may be underestimated. They are
also highly correlated, and our fits do not take the correlation into
account.  Nonetheless, we perform fits out to 13 $h^{-1}$Mpc, keeping
in mind that the errorbars are actually larger due to the
correlations, and that the bias is somewhat degenerate with the choice
of $x_1$.  Based on our investigations, we see no reason to include
mass beyond the virial radius, and so use $x_1=1$ throughout this
work, but larger simulations will be necessary to investigate the
behavior on large scales in more depth.

In addition to correlations of satellite galaxies with the host halo
dark matter, there are correlations between a satellite galaxy and
dark matter particles of its own (sub)halo. Thus, the remaining
uncertainty in modeling satellites is how much of the dark matter
around satellite galaxies remains attached to them.  Analyses of
cosmological simulations show that the typical fraction of mass bound
to subhalos is $\approx 0.1-0.2$, roughly independent of the host halo
mass but strongly dependent on the distance from the host center
\citep[e.g.,][]{1998MNRAS.300..146G,2004MNRAS.tmp..484G,2004astro.ph..5445W}.
Since the fraction of subhalos, $\alpha$, is typically low ($\alpha
\sim 0.2$), the correction due to the subhalo bound mass is small.
The fraction of dark matter remaining is likely to depend on the
details of satellite history, such as the point of closest approach to
the halo center, where the tidal stripping is strongest. It is also
likely to be somewhat correlated with the instantaneous position of
the satellite, but this correlation is not perfect, given the highly
elliptical orbits of satellites observed in simulations
\citep{2000ApJ...544..616G}.  Determining this position in all but the
most massive clusters (where the center is often traced by either
X-ray emission or a cD galaxy) is difficult. For this reason, we will
work with an average quantity rather than some function of radial
distance from the halo center.  We assume the dark matter was tidally
stripped in the outer parts of the halo, but remains unmodified in the
inner parts of the satellite halo. Effectively this means that each
non-central galaxy also has a central contribution, which we model in
the same way as for the central galaxies (i.e., as a halo with mass
$M_{0,i}$ before stripping) out to a fraction of virial radius, but
totally stripped beyond that, yielding g-g lensing signal
$\Delta\Sigma \propto R^{-2}$.
Based on fits (figure~\ref{F:trunc}) we will choose the truncation radius at 
0.4$r_{\rm vir}$, which is equivalent to having 50\%
of the mass stripped, consistent with the average mass loss for subhalos
observed in cosmological simulations \citep{2004MNRAS.tmp..484G,2005ApJ...618..557N}.
We note, however, that given existing measurement errors,
the g-g lensing signal is only weakly sensitive to the
exact value.  Note that this truncation is responsible for the kink in
the stripped satellite profile in the noncentral signal in
figure~\ref{sample}; a more realistic approach might involve smoothing
out this feature.

\subsection{Signal computation}

Using all of the above ingredients, we compute the predicted signal
$\Delta\Sigma(R,M)$ for a given cosmological model and central halo mass
$M_{0}$, central galaxy bias $b(M_{0})$, and $c_g$.  Here we briefly describe the signal
computation for the various contributions.  Note that when we compute
signal in comoving coordinates, the characteristic redshift $z_0$ is
only used when computing the growth factor $b(z_0)$ for use in
normalizing the linear power spectrum, and when determining $M_{\rm nl}(z)$.

First, there is the 1-halo (Poisson) term for the central galaxies,
$\Delta\Sigma_{0}^{P}$.
This term can simply be computed by finding $\Sigma(R)$ and then
$\Delta\Sigma(R)$ for a NFW profile given $M_{0}$ and $c_{\rm dm}$
which is determined from equation~\ref{cm}.  When combining the signal
from central  and satellite galaxies, this term is multiplied by
$1-\alpha$.
There is also a similar term for the satellites,
$\Delta\Sigma_{\rm sat}^{P}$, which
assumes a stripped density profile.  This term is the same as
that for the central galaxies out to $0.4r_{vir}$ (our chosen
truncation radius), then is
proportional to $R^{-2}$ beyond that scale.  When combining signal for
central
and satellite galaxies, this term is multiplied by $\alpha$.

For the remaining terms in the signal, $\Delta\Sigma$ was computed by
finding the galaxy-dark matter cross power spectrum (this approach is described in \cite{2000MNRAS.318..203S}).  The power
spectrum can then be Fourier
transformed to obtain the galaxy-dark matter cross-correlation
function $\xi_{g,dm}$, which yields $\Sigma(R)$
via an integration over the comoving separation 
\begin{equation}
\Sigma(R) = \overline{\rho} \int \xi_{g,dm}[(R^2+\chi^2)^{1/2}]d\chi
\end{equation}
(where we have dropped an unobservable constant term)
and $\Delta\Sigma(R)$ after another integration.  The
power spectrum can be related to the Fourier transform of the halo
profile via $y(k,M)$, where
\begin{equation}\label{ykm}
y(k,M) = \frac{\tilde{\rho}(k,M)}{M} = \frac{1}{M}\int_{0}^{x_{1}r_{vir}} 4\pi r^2 dr
\rho(r,M) \frac{\sin{(kr)}}{kr}
\end{equation}
We can define $y_{\rm g}$, the Fourier transform of the radial profile of
galaxies using a NFW profile with concentration $c_g$,
and $y_{\rm dm}$, the Fourier transform of the halo dark matter
profile with concentration
$c_{\rm dm}$.

 This
procedure was used to get the noncentral Poisson term $\Delta\Sigma_{\rm nc}^{P}$,
 which is obtained from
\begin{equation}
P_{\rm nc}^{P}(k) = \frac{1}{(2\pi)^3 \overline{n}} \int f(\nu)d\nu
\langle N(M)\rangle y_{\rm dm}(k,M) y_{\rm g}(k,M)
\end{equation}
and describes the correlations between satellite galaxies and their
host halo dark matter.
Because this is a noncentral term, it must be multiplied by $\alpha$
when computing its contribution to $\Delta\Sigma$.

This procedure was also used to compute the central and noncentral
halo-halo terms, $\Delta\Sigma_{\rm cent}^{hh}$ and
$\Delta\Sigma_{\rm nc}^{hh}$, which dominate on large
scales.  
The power spectrum for the central halo-halo term (describing correlations between the central galaxy of one host halo and the
dark matter of another host halo), is
\begin{equation}
P_{\rm cent}^{hh}(k) = b(M_{0})P_{lin}(k) \int f(\nu)d\nu
\,b(\nu) y_{\rm dm}(k,M)
\end{equation}
where the integration over $\nu$ includes all values.  
Note that dark matter must on
average not be biased,
and therefore the last term for a general $b(\nu)$ must
reduce to 1 on large scales where $y_k=1$.  For the bias
expression used in this paper, equation~\ref{bias}, the average bias is 1.06 rather than 1,
but in order to reproduce the relation $P_{\rm cent}^{hh}(k) =
b(M_{0}) P_{lin}(k)$ on large scales, we have artificially
normalized the integral to 1 on large scales.
This central halo-halo power spectrum must be multiplied by $(1-\alpha)$ when combining the signal for
a mixture of central and noncentral galaxies.  

For the remaining noncentral
halo-halo term, we have
\begin{multline}
P_{\rm nc}^{hh}(k) = P_{lin}(k) \int f(\nu)d\nu \,b(\nu) y_{\rm
  dm}(k,M) \times \\
  \frac{\overline{\rho}}{\overline{n}}\int f(\nu) d\nu \, b(\nu) \frac{\langle N(M)\rangle}{M} y_g (k,M),
\end{multline}
where the first integral, as above, must be 1 on large
scales, and the second term can be normalized via
\[
\frac{\overline{n}}{\overline{\rho}} = \int f(\nu)d\nu\, \frac{\langle N(M)\rangle}{M}
\]
and (once normalized) equals the average bias $\langle b \rangle$ for satellite
galaxies of this mass.  Consequently, $P_{\rm nc}^{hh}(k)$ reduces to
$\langle b \rangle P_{lin}(k)$ on large scales, and must be multiplied by $\alpha$ when combining signal for central
and satellite galaxies.  When the central and noncentral signal are
combined, we then have on large scales an effective bias $b_{\rm eff} =
(1-\alpha)b(M_{0}) + \alpha\langle b\rangle$.  

\begin{figure}
\epsfig{file=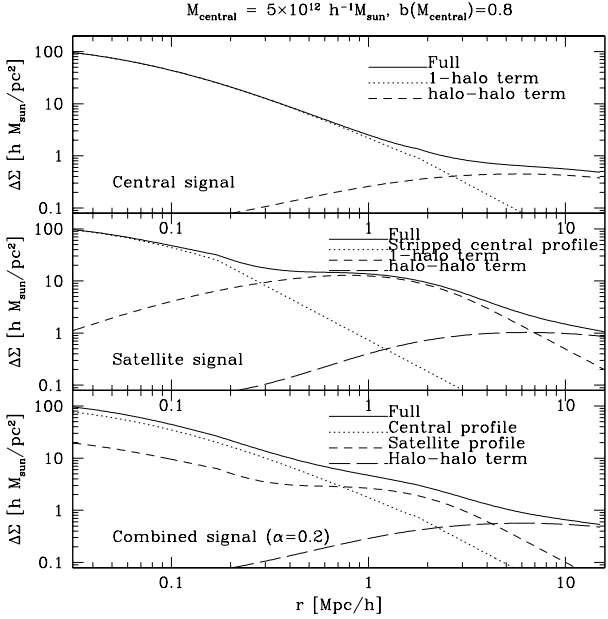,  height=\linewidth, angle=0}
\caption{Halo model predictions for $\Delta \Sigma(r)$ for a central
  halo mass of $M_{0}=5\times 10^{12} h^{-1}M_{\sun}$.  The top
  panel shows the central signal with the 1-halo and halo-halo terms
  shown separately, assuming $b(M_{0})=0.8$.  The middle panel
  shows the noncentral signal, with the stripped satellite central profile, the
  1-halo noncentral term, and the halo-halo noncentral term shown
  separately as labeled. The bottom panel shows the combined signal,
  with satellite fraction $\alpha=0.2$.} 
\label{sample} 
\end{figure}

To illustrate the effects of the various contributions to the signal
discussed in this section, figure~\ref{sample} shows the predicted
central (top), noncentral (middle) and total (bottom) signal for $\Omega_m=0.3$,
$\sigma_8=0.9$, $b(M_{0})=0.8$, and $M_{0} = 5\times 10^{12} h^{-1}M_{\sun}$.
The central signal is the sum of two contributions:
\begin{equation}
\Delta\Sigma_{\rm cent} = \Delta\Sigma_{\rm cent}^{P} +
\Delta\Sigma_{\rm cent}^{hh}
\end{equation}
and the satellite signal is the sum of three contributions:
\begin{equation}
\Delta\Sigma_{\rm sat} = \Delta\Sigma_{\rm sat}^{P} +
\Delta\Sigma_{\rm nc}^{P} + \Delta\Sigma_{\rm nc}^{hh}.
\end{equation}
For the combined signal we used satellite fraction $\alpha=0.2$. 

\section{Halo model versus simulations}

As described in the introduction, the 
main motivation for 
the comparison between the halo model and simulations 
is that 
the halo model makes several approximations,
so it must be verified and calibrated on realistic simulations. 
We may, of course, find that the simulations reproduce the data 
perfectly, in which case
one could argue that the halo model itself is unnecessary. However, there 
may be different cosmological models or different treatments of how 
to populate halos and subhalos with galaxies 
that all agree equally well with the
data. Rather than identify them with many expensive numerical 
simulations, one can use the halo model to do the same. 

As discussed in \S2, the simulations used here cannot reliably resolve low mass
halos, so we restrict our analysis to 3 luminosity bins, each 1 magnitude
wide in $r$-band, L3 $[-19,-20]$,
L4 $[-20,-21]$ and L5 $[-21,-22]$, using the notation from \cite{2005PhRvD..71d3511S}. 
We do not use the brightest  bin (L6 $[-22,-23]$), since most of these 
galaxies reside in clusters, and our simulation box is too small 
to properly sample these. 

To test how well g-g lensing analysis can extract various halo model 
parameters, we take the full g-g lensing signal from simulations and 
fit for 4 parameters: $M_{0}$, $b(M_{0})$, satellite fraction 
$\alpha$, and radial distribution of satellites $c_g$.
When performing the fits to the central galaxy signal alone, we also 
explored the sensitivity to the 
dark matter halo concentration parameter $c_{\rm dm}$.
We assign errors to the radial points using the actual errors that we 
find in SDSS analysis \citep{2005PhRvD..71d3511S}. As the survey continues to take data, 
the error amplitude will decrease, but the 
relative errors will remain the same, so one can simply rescale the errors by the appropriate amount.  
Note, however, that a deeper survey, such as RCS \citep{2004ApJ...606...67H}, can provide 
better information on g-g lensing at small radii, which would be 
particularly useful for determination of the halo concentration parameter 
$c_{\rm dm}$. 

\begin{figure}
\epsfig{file=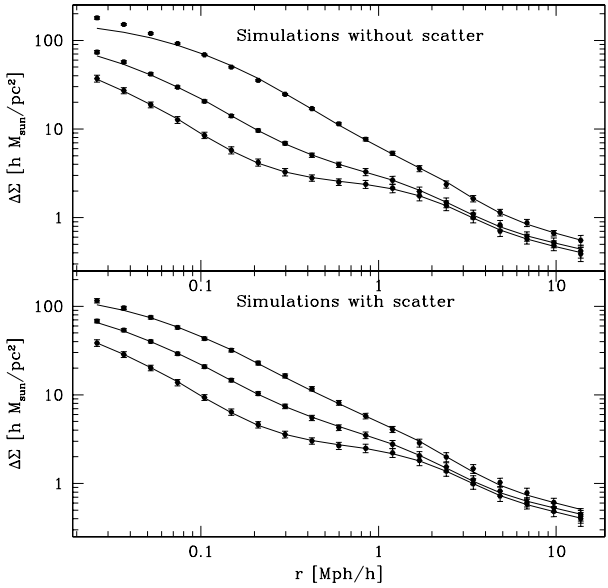,  height=\linewidth, angle=0}
\caption{$\Delta \Sigma(r)$ for 
L3, L4, and L5 luminosity bins, bottom to top. The top panel is 
for simulations without scatter between $V_{\rm max}$ and $L_r$, 
while the bottom panel is with scatter. 
Points are the simulation data with errors taken from observations
and scaled down by a factor of 10, with simulation errors added in
quadrature. The solid line is the halo model fit. Resolution effects
become important below  r$\simeq$50~$h^{-1}$kpc. Thus, for the fits
we used   simulation data  above this scale only; the disagreement 
for the inner few points is caused by insufficient resolution in simulations.} 
\label{fig1} 
\end{figure}

Figure \ref{fig1} shows the lensing signal $\Delta \Sigma(R)$ as a function 
of transverse separation $R$ from the simulations, together with the
halo model signal 
that fits the simulations best. We show the results separately for 
all luminosity bins with and without scatter.
The fits are done using SDSS error bars \citep{2005PhRvD..71d3511S}
reduced by a factor of 10, with the error from simulations added in quadrature.

To obtain an estimate of the minimum reliable scale in simulations, 
we performed a convergence test
with respect to $\Delta \Sigma$, using the low and high resolution (8 times more particles)
simulations of cluster CL2 of \cite{tasitsiomi_etal04}.   
We found that a reliable minimum scale is $\sim$50~$h^{-1}$kpc
and hence for the fits we only use simulation data on scales larger than that. 
In fact,  resolution can affect both the dark matter density
profiles, making them somewhat shallower than they actually are,
and the number of subhalos in the innermost host halo regions. 

The most striking feature of the figure is how well the halo model
reproduces the simulations with 3 free parameters (central halo mass, 
satellite fraction and satellite radial distribution, assuming that bias 
is determined by the halo mass, which is confirmed when we compare the 
fitted values to those predicted). 
As discussed, the small disagreement on small scales is actually caused by finite 
resolution in simulations, which leads to an enhancement of 
$\Delta\Sigma$. 

Figure \ref{fig1} shows the signal from simulations for a combined sample of 
central and satellite galaxies, such as would be observed
in a typical 
luminosity-selected sample with no additional selections applied. 
We have also done the fits separately for central and noncentral 
components. 
If one takes galaxies in dense 
regions, then the fraction of satellites is likely to be increased, while 
if one takes the sample of galaxies in low density regions, the 
satellite fraction is likely to be decreased. Similar effect can 
be achieved by color selection; for example, when taking blue
galaxies, the satellite fraction is likely to be decreased. 
These two sub-samples therefore
provide the extreme cases of such selection, and are useful as an 
indication of what additional information one may be able to extract 
from the g-g lensing data given such selections. 

\subsection{Central galaxies}

\begin{figure}
\epsfig{file=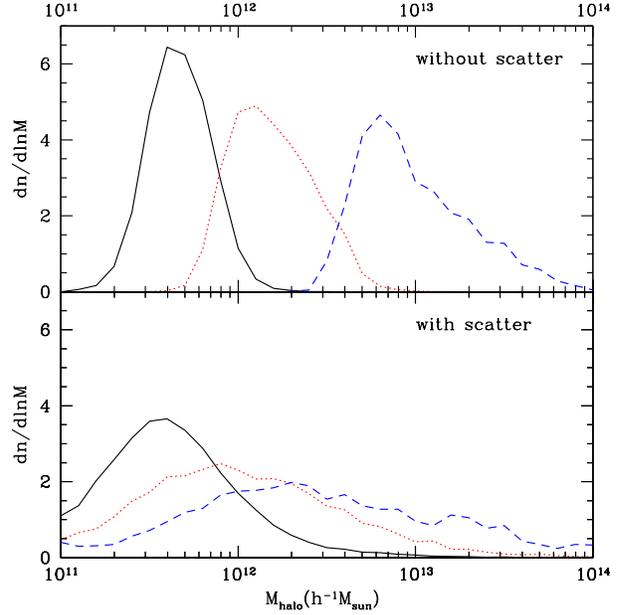,  height=\linewidth, angle=0}
\caption{Halo mass distribution for central galaxies for 
L3 (solid), L4 (dotted), and L5 (dashed) luminosity bins. The top panel is 
for simulations
without scatter, the bottom is with scatter. Median and mean masses
are given in table 1.
\label{fig:mhalo} }
\end{figure}

There are several aspects of the halo model that we wish to test. 
We begin with 
the halo mass distribution for central galaxies. 
For each of the magnitude bins we 
have a distribution of central halo masses from the simulations, 
as shown in figure 
\ref{fig:mhalo}. 
As is clear from this 
figure, the halo mass distributions are reasonably 
narrow in the simulations without scatter, while in the case with scatter they 
are significantly broader, with the width and assymetry of the distribution
increasing for higher luminosity bins. This effect is also evident from the difference between 
the mean and the median of the mass distributions 
given in table \ref{mhalo}, which reaches a factor of 5 in the 
brightest bin with scatter. The width is largest for the 
brightest bin because the galaxies
are on the exponential slope of the luminosity function, while many of
the associated halos are on the exponential tail of the halo mass function. 
In light of the large width, it is difficult to speak of a single mass associated
with a given luminosity bin, yet we attempt to determine a single mass
using the fitting rather than the full shapes of the curves in
Fig.~\ref{fig:mhalo}, so we must investigate further to understand the
consequence of this choice. The relevant questions associated with
the halo model fitting are: What is the meaning of the best fit halo mass 
from the halo model
fits?  How much of an error is one making by using the best-fit
masses if one is interested in the 
mean (median) halo mass for a given sample? Can one identify from 
the lensing signal itself the width of the central halo mass distribution? 

To address the first two questions, we can compare
the mean and median of the distribution
to the best-fit masses using a NFW 
profile. We perform NFW fits in two 
separate ways. First, we fit the 
central galaxy component signal only to a  NFW profile, using all of the information out to 
260$h^{-1}$kpc. We chose this scale motivated by  observational studies where this scale is used
to define the aperture mass \citep[e.g.,][]{2001astro.ph..8013M}. We fit simultaneously for concentration and virial mass. 
In the second case, we use the full combined signal, effectively 
cutting out the information from the central halos at large transverse separations, since the signal 
there is dominated by satellites sitting in more massive halos, whose 
signal dominates at larger radii. In this case we assume concentration
$c_{\rm dm}$
based on equation \ref{cm}, and also fit simultaneously for $b(M_{0})$, $c_g$ and $\alpha$. Both fits give similar results for virial masses 
(table \ref{mhalo}).

If the central galaxy halo mass distribution is narrow, the halo model 
is able to determine it quite accurately, as shown in table \ref{mhalo}. 
If the halo mass distribution is broad, then there is no typical mass, and 
in general the g-g lensing mass
determination from NFW fits falls 
between the mean and the median mass of the halo population. 
If one is interested in the mean halo mass, then the NFW fits underestimate 
the mass by 10-30\% at L3 and 10-55\% at L4 and L5, so one must 
increase the best-fit NFW masses by up to a factor of 2  or even more 
at the bright end when 
scatter is significant. If the median masses
are of more interest, then one must decrease NFW masses by similar amounts. 
At low luminosities,
these corrections are small, but become increasingly important for 
brighter galaxies because the halo mass distribution is broader. 
Even more problematic is the fact that it is 
difficult to assess the corrections from the fits itself, since the 
fits are good in both cases, but the corrections vary from essentially 
none in the case without scatter to over a
factor of two in the case of scatter. 
To improve the reliability of the mass determination, one must therefore
use samples of galaxies with as little scatter as possible. 
Below we discuss this point further. 

\begin{table*}
\caption{Mean and median halo masses for central galaxies 
in simulations with and without 
scatter for 3 luminosity bins. 
Also given are the mass and concentration from NFW profile fits to the
g-g lensing signal for the
central component only, and using the full (central and satellite) 
g-g lensing signal. 
For concentration $c_{\rm dm}$, we only show the values for the fits to the central 
component; we assume equation~\ref{cm} for the fit to the full
signal.}
\begin{center}
\small
\begin{tabular}{ccccccc}
\hline
Luminosity &
$\langle M \rangle_{\rm central}$&
$M_{\rm median, central}$  &
$M_{\rm NFW fit, central}$  &
$c_{\rm NFW fit, central}$  &
$M_{\rm NFW fit, full} $  &
Scatter in $V_{\rm max}-L$ 
\\
\hline
&
$10^{12}h^{-1}\rm\ M_{\odot}$&
$10^{12}h^{-1}\rm\ M_{\odot}$ & 
$10^{12}h^{-1}\rm\ M_{\odot}$ &
&
$10^{12}h^{-1}\rm\ M_{\odot}$  
\\
\hline
\\
L3 [-19,-20] & 0.506  & 0.460  & 0.476 & 12.8 & $0.42\pm 0.03$ & no \\
L4 [-20,-21] & 1.79  & 1.47  & 1.61 & 12.7 & $1.52\pm 0.07$ & no \\
L5 [-21,-22]  & 13.2  & 8.36  & 9.71 & 12.2 & $12.6\pm 0.3$ & no\\
L3 [-19,-20] & 0.745  & 0.390  & 0.579 & 11.6 & $0.47\pm 0.04$ & yes \\
L4 [-20,-21] & 2.91  & 1.10  & 1.86 & 10.9 & $1.42\pm 0.09$ & yes \\
L5 [-21,-22]  & 11.7  & 2.34  & 6.21 & 10.3 & $4.8\pm 0.2$ & yes \\ 
\end{tabular}
\end{center}
\label{mhalo}
\end{table*}

\begin{table*}
\caption{
Fit results from a fit for central halo mass and bias $b(M_{0})$ to the signal for central galaxies, using information out
to $13h^{-1}$Mpc. Because neighboring bins are correlated on large scales, the bias
errors are likely to be underestimated. The true bias values are from \protect\cite{2004MNRAS.355..129S}.}
\begin{center}
\small
\begin{tabular}{ccccc}
\hline
Luminosity &
$M_{\rm NFW fit, central}$  &
$b(M_{\rm central})$ & $b_{\rm true}$ & Scatter in $V_{\rm max}-L$
\\
\hline
&
$10^{12}h^{-1}\rm\ M_{\odot}$& &

\\
\hline
\\
L3 [-19,-20] & $0.45\pm 0.02$ & $0.65\pm 0.04$ & 0.68 & no \\
L4 [-20,-21] & $1.62\pm 0.03$ & $0.68\pm 0.04$ & 0.72 & no \\
L5 [-21,-22] & $11.8\pm 0.2$  & $0.90\pm 0.05$ & 1.00 & no \\
L3 [-19,-20] &    $0.54\pm 0.02$ & $0.64\pm 0.04$ & 0.68 & yes \\
L4 [-20,-21] &    $1.82\pm 0.05$ & $0.73\pm 0.04$ & 0.72 & yes \\
L5 [-21,-22] &   $6.8\pm 0.2$   & $1.00\pm 0.06$ & 0.93 & yes\\
\end{tabular}
\end{center}
\label{mhalobias}
\end{table*}

Despite the wide distribution of halo masses, the NFW fitting is able
to extract the mean concentration parameters of the dark matter halos
correctly.  This is particularly true for the lower luminosity cases
L3 and L4, where equation \ref{cm} with $M_{\rm nl}=9\times
10^{12}h^{-1}M_{\sun}$ ($\Omega_m = 0.3$, $\sigma_8=0.9$) gives
$c_{\rm dm}=12-14$ for the mean virial mass in these bins, in good
agreement with the fitted value $c_{\rm dm}\sim 13$.  In L5 without
scatter, the fitted value, $c_{\rm dm}\sim 12$, is somewhat higher
than the predicted value $c_{\rm dm}\sim 10$.  It is possible that for
this bin, the difference in best-fit masses between the two kinds of
fits stems from the assumption of a lower concentration in the fit to
the full signal, which would encourage a higher best-fit mass, as is
observed.  We note here, however, that the NFW dark matter profile is
not a perfect fit to the simulations, so there are small differences
in the NFW fits to 3-d data versus 2-d $\Delta \Sigma$ data. Hence,
these differences may be a reflection of the uncertainties in NFW
parameterization rather than a failure of the halo model.

Adding scatter leads to a modest decrease in the fitted concentration 
parameters. This effect is relatively small in the faintest bin and 
increases towards the brighter bins, where scatter has more effect. 
Thus, averaging over 1 magnitude bins in luminosity should allow
one to extract the concentration parameter from g-g lensing signal
with little or no bias for galaxies below $L_*$. 
This is particularly the case if one can identify a galaxy population with a
narrow distribution of halo masses. An example of this would be
selecting elliptical galaxies that lie in a narrow strip on the 
fundamental plane which corresponds to the dynamical mass
\citep{2004NewA....9..329P}.  
Another example are galaxies selected by stellar mass
\citep{2003MNRAS.341...33K}, which is likely  
to be a more faithful tracer of dynamical mass than the luminosity
itself.  Careful selection of the lens population in one of these
ways, via a reliable tracer of the dynamical mass, should help ensure
that the best-fit central halo mass has a simple interpretation. 

If one assumes a given cosmological model, then the concentration parameter 
is fixed, and the difference between the predicted and observed 
concentration can be used to study the width of the halo mass distribution. 
We see that even in the cases of a very broad distribution, as in L5, 
the differences in the average concentration parameter are modest and it will be 
difficult to observe them directly in observations. Moreover, 
one can change concentration for a given halo mass
by changing the cosmological parameters (primarily nonlinear mass), so the two 
effects are, to some extent, degenerate. 

Table~\ref{mhalobias} shows results for fits to the central signal using all scales (out to
$13h^{-1}$Mpc) to determine $M_{0}$ and bias $b(M_{0})$.  This determination of $b$ is particularly useful since on
large scales the central signal is completely negligible compared to
the halo-halo term, so the bias is easily extracted from the signal
amplitude. However, the errorbars are likely underestimated due to the
correlations in the correlation function bins.  As shown in
table~\ref{mhalobias}, the bias values are reasonable for central
halos with those masses when compared against the results in~\cite{2004MNRAS.355..129S}.

\begin{figure}
\epsfig{file=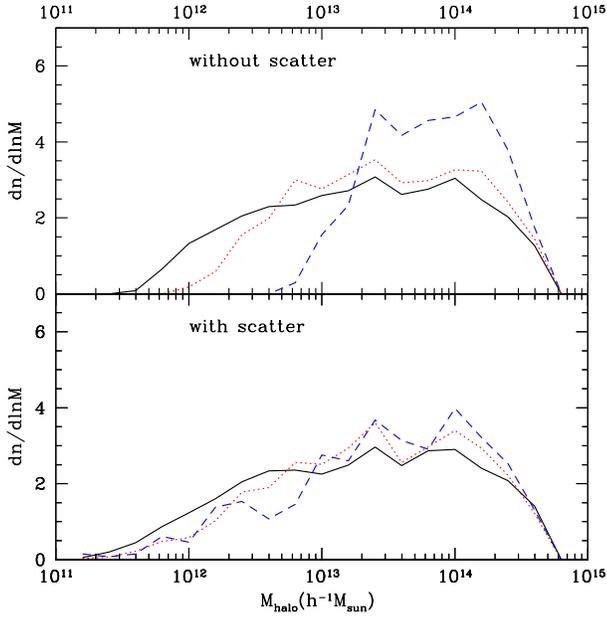,  height=\linewidth, angle=0}
\caption{Halo mass distribution for hosts of satellite galaxies for 
L3 (solid), L4 (dotted), L5 (dashed) luminosity bins. The top panel 
shows the case where there is no scatter between $V_{\rm max}$ and $L_r$, 
while the bottom panel assumes scatter, as described in \S~\ref{sec:sims}. Note that scatter makes the 
distributions nearly independent of luminosity. 
\label{fig:mhalosat} }
\end{figure}

\subsection{Satellites} 

We turn next to the satellite signal. We begin with the distribution of 
host halo masses for satellites in a given narrow luminosity range. 
These are very broad, as shown in figure \ref{fig:mhalosat}. 
This is expected, since a satellite of a given luminosity can be part of 
a small group or a large cluster and there is little relation between 
the halo mass and satellite luminosity. 
The latter is particularly true for the case with scatter, where the 
differences between the distributions are very small and only appear at 
the low mass end of the host halo masses. 

\begin{figure}
\epsfig{file=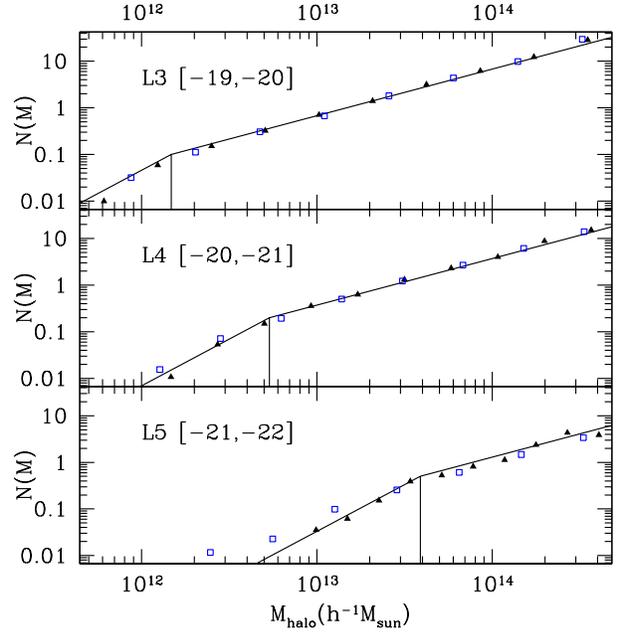,  height=\linewidth, angle=0}
\caption{Mean halo occupation number as a function of 
halo mass for satellite galaxies for 
the three luminosity bins, as labelled. Solid triangles are 
without scatter, open squares are with scatter. The solid line is the 
analytic approximation used here for the halo model, the vertical line delineates
the break at $3M_{0}$ as used in GS02.
\label{fig:nm}}
\end{figure}

Instead of discussing the actual mass distribution we can
phrase it in terms of the average halo occupation model, $\langle N(M)\rangle$. To obtain 
the actual halo mass distribution of satellites, one has to multiply $\langle N(M)\rangle$
with the halo mass function (equation \ref{fnu}). The latter is  
well determined from simulations 
for masses in the range $(10^{12}-10^{14})h^{-1}M_{\sun}$, the range 
of interest here \citep{1999MNRAS.308..119S,2001MNRAS.321..372J}. 
The halo occupation $\langle N(M)\rangle$ as a function of halo mass $M$ is shown in figure 
\ref{fig:nm} for the 3 magnitude bins. We show simulation results without
scatter and with scatter. 
We see that in all cases $\langle N(M)\rangle\propto M$ (i.e., $\epsilon=1$)
is a good fit to the
simulations above $3M_{0}$, while below this mass the number of 
satellites declines more rapidly. The distributions with and without 
scatter are very similar for L3 and L4, while for L5,
the scatter increases the abundance of galaxies at the low mass end. 
We also show
the simple step function model of GS02 and an improved model where the 
decline below $3M_{0}$ is more gradual, modeled here as
$\langle N(M)\rangle\propto M^2$ below $3M_{0}$ (the improved
model was used for all fits).
Note that the overall amplitude is not relevant in these relations, 
since we normalize 
it to the overall fraction of the 
galaxies that are satellites in any given sample. 

\begin{table*}
\caption{Satellite fraction and concentration parameter  
of satellite radial distribution from simulations and from 
the halo model fits for fits out to 2$h^{-1}$Mpc without 
h-h term and out to 13$h^{-1}{\rm Mpc}$ including 
the h-h term. 
}
\begin{center}
\small
\begin{tabular}{cccccccc}
\hline
Luminosity &
$\alpha_{\rm true}$&
$\alpha_{\rm fit, 2h^{-1}{\rm Mpc}}$  &
$\alpha_{\rm fit, 13h^{-1}{\rm Mpc}}$  &
$c_{g, \rm true}$  &
$c_{g, \rm fit,2h^{-1}{\rm Mpc}}$  &
$c_{g, \rm fit,13h^{-1}{\rm Mpc}}$ &
Scatter in $V_{\rm max}-L$   
\\
\hline
\\
L3 [-19,-20] & 0.23  & $0.26\pm 0.03$ & $0.22\pm 0.02$  & 1.5 & $2.2\pm 1.1$ & $3.2\pm 0.4$ & no \\
L4 [-20,-21] & 0.20 & $0.21\pm 0.01$ & $0.18\pm 0.02$ & 3.0 & $4.7\pm 1.8$ & $5.9\pm 0.7$ & no\\
L5 [-21,-22]  & 0.15 & $0.15\pm 0.02$ & $0.10\pm 0.02$ & 3.5 & $1.5\pm 1.3$ & $2.1\pm 0.4$ & no  \\
L3 [-19,-20] & 0.23 & $0.26\pm 0.03$ & $0.22\pm 0.02$  & 1.5 & $2.8\pm 1.4$ & $3.9\pm 0.5$ & yes \\
L4 [-20,-21] & 0.21 & $0.20\pm 0.02$ & $0.17\pm 0.02$  & 2.5 &  $10\pm 4$ & $12\pm 1$ & yes \\
L5 [-21,-22]  & 0.19 & $0.16\pm 0.05$ & $0.16\pm 0.03$ & 2.3 & 20 & 20 & yes \\
\end{tabular}
\end{center}
\label{alphacg}
\end{table*}
  
Examination of figure~\ref{fig1} on $\sim 1$ Mpc scale shows that 
scatter has little effect on the noncentral signal in L3 and L4, while for L5 the effect is much more 
significant: scatter increases the abundance of low mass halos and so 
reduces the g-g lensing signal. 
Despite this effect, we are able to reproduce the 
fraction of satellites in the overall sample rather well: 
as shown in table 
\ref{alphacg}, the best-fit satellite fractions are in most cases 
within 10\% of
the correct value. 
This quantity therefore appears to be 
relatively robustly extracted from the data.  
This works best if we simply ignore the h-h term and only fit out to 
2$h^{-1}$Mpc. In a more general fit with the h-h term and out to larger 
radii, we find there is some degeneracy between the components, and the 
extracted satellite fraction can be too low in some cases 
(table \ref{alphacg}). 

For the halo model, we have to choose the radial distribution of
galaxies, which we model as a  NFW profile with its own concentration
parameter $c_g$. The concentration of subhalo radial distribution
depends sensitively on how the subhalos are selected.
\cite{2005ApJ...618..557N} show that selection of subhalos by their
maximum circular velocity $V_{\rm max}$ results in concentrations of
subhalo number density profiles in clusters in the range $\sim 1.5-4$,
while selection using luminosity or stellar mass results in
considerably larger concentrations ($c\sim 3-10$). Although
luminosities have been assigned to the subhalos in our catalogs, the
assignment did not take into account the position of the subhalo within
its host and depended only on the subhalo's $V_{\rm max}$.  We can
therefore expect that the radial distribution of the cluster subhalos
in simulations will be characterized by  concentrations similar to
those measured for subhalos selected using $V_{\rm max}$ (i.e.,
$c_g\sim 1.5-4$).  Results of our analysis over a broad range of host
halo masses, between $10^{12}h^{-1}M_{\sun}$ to
$10^{14}h^{-1}M_{\sun}$, indeed suggest values of $c_g$
significantly lower than $c_{\rm dm}$. Table \ref{alphacg} gives the
true $c_g$ value from simulations weighted over the halo mass
distribution for a given luminosity bin.

It should be noted that shallow profiles for the satellite
distribution are not a unique prediction of the simulations.
Semi-analytic simulations find $c_g \sim c_{\rm dm}$ by identifying
galaxies before merging into larger halos and following their identity
even after their surrounding dark matter has been stripped off
\citep{2004MNRAS.352L...1G}.  Observations suggest that the galaxy
distribution can be well described by the NFW profile in rich clusters
\citep[e.g.,][]{1997ApJ...478..462C,2004ApJ...610..745L,
  2004astro.ph..10467H} These works also find that galaxies have
concentrations significantly lower than expected for their parent
halos.  Since the actual value of $c_g$ depends on the physical
processes that affect the specific type of galaxy under consideration,
we treat $c_g$ as a free parameter to be determined by fitting the
simulations (or data).

We find that in order to fit the simulations, we need the satellite distribution 
to be shallow (table \ref{alphacg}), consistent with 
the value found by directly fitting to the satellite distribution in the same simulations, except for the two highest luminosity bins. 
As discussed above,
the low value of $c_g$ found in simulations 
does not imply that the same value has to be appropriate to 
fit the data: the radial distribution of galaxies of a given luminosity 
inside larger halos is affected by star formation and dynamical 
processes both before and after falling into the larger halos, 
which is likely to be quite complicated.  
What is important for the current discussion is that the halo model 
should reliably determine the radial distribution of galaxies from the 
g-g signal, allowing one to determine it from
observations. 

Table~\ref{alphacg} also shows the best-fit value of $c_g$ from the
full signal.
As shown, even with the ten times reduced SDSS errorbars, we are not highly
sensitive to $c_g$.  
The result for L5 with scatter is quoted
without errorbars because the fit program chose the maximum value
allowed in the fits, with very small errorbars.  This result (and the
high value for L4 with scatter)
may be an
indication that $\Delta\Sigma$ averaged over such a large range of
halo masses does not lend itself well to this type of fitting for a
single value of $M_{0}$ and $c_g$.  Once
again, this result emphasizes that the fitting is most meaningful for
groups of lens galaxies chosen to have a narrow distribution in
central halo mass, which may not necessarily be the case for $r$-band
luminosities even with narrow (1-magnitude wide) bins.
The results of the fits
suggest that the radial distribution can in principle 
be extracted from the 
observations, although we note that when applying this to the actual 
data, the observational errors were found to be large and 
unable to provide a strong constraint from the current 
data samples \citep{2005PhRvD..71d3511S}.  Note that when the fit was
performed to $13 h^{-1}$Mpc using the combined signal, the bias values $b(M_{0})$
were found to be lower than the values in table~\ref{mhalobias}, and
an examination of figure~\ref{sample} suggests that on intermediate
scales (2--4 $h^{-1}$Mpc), there is some interplay between $c_g$ and
$b(M_{0})$ that is causing the high best-fit values of $c_g$ and
the low values of $b(M_{0})$.  A better approach may be to
determine the bias $b(M_{0})$ using the central signal alone as in table~\ref{mhalobias},
then fix it to that value in the fits for the full signal rather than
allowing it to vary; this will encourage lower, more reasonable values
of $c_g$.

We note here that the satellite radial distribution is more important for the 
overall shape of the satellite contribution to g-g lensing signal than 
the halo occupation number as a function of mass $\langle N(M)\rangle$.
Changes in the slope $\epsilon$ in
$\langle N(M)\rangle\propto M^{\epsilon}$ mostly change the amplitude of the  
satellite signal without changing its shape (see figure 7 in GS02). 
This result is expected, since the satellite signal is dominated by halos 
with masses in the range $(10^{13}-10^{14})h^{-1}M_{\sun}$. Change in the 
slope changes their relative abundance, but more or less preserves the radial 
shape of their signal. Consequently, when we attempted fits for both
$\epsilon$ and $\alpha$, the slope $\epsilon$
was almost completely degenerate with 
the fraction of satellites $\alpha$, but the actual halo mass 
probability distribution in this mass range was less affected.
Hence, as discussed already, we did not fit for $\epsilon$, but instead
assumed fixed $\epsilon=1$ throughout. 

\begin{figure}
\epsfig{file=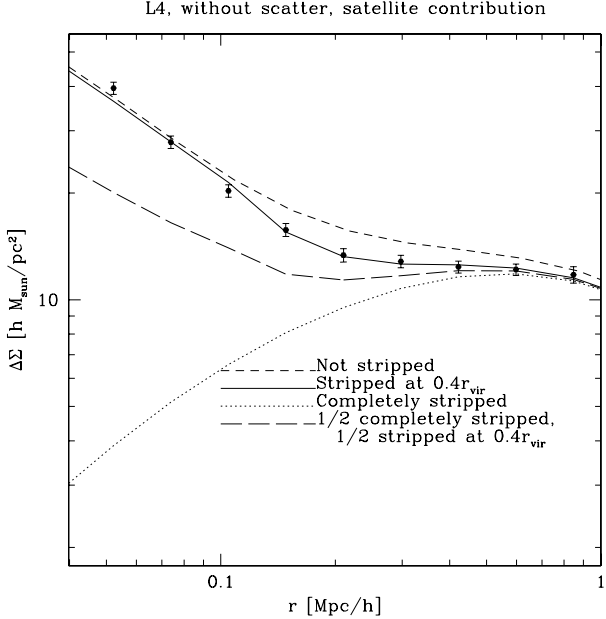,  height=\linewidth, angle=0}
\caption{$\Delta \Sigma(r)$ for 
satellite contribution to $\Delta\Sigma$ from L4 without scatter.  The
points are from simulations, and the lines are (as labeled) the
best-fit signal with no mass stripped, partially stripped (truncation
at $0.4r_{vir}$), fully stripped, and a mixture of half partially
stripped and half fully stripped.}
\label{F:trunc} 
\end{figure}

Another aspect of the halo model that can be tested with simulations is 
the amount of dark matter attached to subhalos within the larger halo,
though as discussed, we only approach this problem in an average sense. 
For g-g lensing, the tidal stripping of the outer layers of dark matter attached 
to the satellite is of little importance, since the lensing signal of that 
component is swamped by the host halo dark matter signal. 
In GS02 it was simply assumed that all of the 
dark matter mass is attached to the 
subhalo, so that in the inner region, the lensing signal of the 
satellites is the same as of the central galaxies of the 
same luminosity. Figure~\ref{F:trunc}, which shows the noncentral
signal for one luminosity bin, with the best-fit signal assuming 
unstripped, moderately stripped, and completely stripped subhalos, shows that this is a good approximation 
in the inner parts. 
One should perhaps not be too surprised by this agreement, since by 
construction in these dissipationless simulations, the 
galaxies sit on top of subhalos that have survived all the merging and 
tidal stripping inside the halos. Reality could be more complicated: one 
could have dark matter in subhalos entirely stripped while the 
more compact stellar component is preserved. 

While the agreement between the simulations and halo model is already 
good, one can improve the agreement further by modeling 
the transition between the satellite lensing signal 
and the host lensing signal, which occurs around 100-200 $h^{-1}$kpc. 
We assume that only the dark matter within 
0.4 of original 
virial radius is attached to the satellite (figure~\ref{F:trunc}). This 
requires that on average 50\% of the dark matter is stripped from 
the satellites. 
We find the signal to be relatively insensitive to the 
exact value of truncation radius and consequently g-g lensing cannot 
be used as a strong probe of tidal stripping process; however, as shown in
figure~\ref{F:trunc}, we can exclude the extreme possibility that all halos are fully
stripped.
To study this issue observationally, it is best to select a galaxy sample 
with a particularly high satellite fraction $\alpha$, such as the 
galaxies in dense enviroments. 

\section{Conclusions}

The goal of this paper is to present a detailed comparison of halo
model and simulation predictions of galaxy-mass correlations as
observed in weak lensing.  In its simplest form, and for a given
cosmological model, the halo model has 3 free parameters, which depend
on the physics of galaxy formation.  These are the typical halo mass
of central galaxies (where ``typical'' is between the median and mean
halo mass), the fraction of galaxies that are satellites, and the
radial distribution of satellites inside the larger halo.

The halo model provides a fairly accurate description of the g-g
lensing signal from simulations, in the sense that it can determine
the virial mass distribution of central and non-central galaxies and
their relative ratios, as well as the radial distribution of both
galaxies and dark matter.  One must be careful when interpreting the
halo masses from NFW fits if the halo mass distribution is broad as is
expected for luminous galaxies ($L>$ few $L_{\ast}$). In this case,
g-g lensing is determining something between the mean and the median
mass and may not recover the correct radial distribution of the
satellite galaxies in their host halos.  If one is interested in the
mean halo mass, as in the case of application to bias studies
\citep{2005PhRvD..71d3511S}, then an upward correction needs to be
applied; this correction is small at the faint end but can be quite
significant at the bright end, especially if there is a significant
scatter in the luminosity-halo mass relation.

There are additional parameters that can be added to the description,
such as the generalization of the assumed halo mass probability
distribution, but they do not improve the fit and are strongly
correlated with these 3 parameters.  Galaxy-galaxy lensing also allows
a determination of the halo mass profile of both central galaxies and
satellites. In the latter case, one can constrain the amount of
satellite mass stripping inside larger halos, which determined how
much dark matter remains attached to the satellites.  To investigate
the radial profiles of dark matter it is best if one identifies
samples that consist predominantly of central or satellite galaxies,
respectively.

Galaxy-galaxy lensing is also quite sensitive to the dark matter concentration 
parameter $c_{\rm dm}$, which (in standard cold dark matter 
models) is not a free parameter but is 
fixed for any given cosmological model. 
It varies modestly as a function of cosmology, 
so in principle one could use this feature to determine the 
cosmological model. For example, for a typical $L_*$ 
galaxy with mass 
around $(1-2)\times 10^{12}h^{-1}M_{\sun}$, we have 
$c_{\rm dm}=12$ in a $\sigma_8=0.9$, $\Omega_m=0.3$ cosmology, which 
drops to $c_{\rm dm}=10$ if $\sigma_8=0.7$ is used instead.
This change is small and we find that it is 
degenerate with the amount of 
scatter, which also leads to a lower value of concentration 
parameter. With current samples, one cannot yet reliably distinguish 
between these models, but one can test the overall consistency.
The accuracy and reliability
with which the concentration can be extracted from the data 
may be improved if galaxies in underdense regions, for which 
the satellite fraction is lower, are selected. In this case, 
one can use the signal to a larger distance since it is not swamped 
by the noncentral component. Concentration can also be determined 
more accurately with deeper surveys, which can probe dark matter 
halo at smaller separations \citep{2004ApJ...606...67H}, although modeling angular 
projection effects in the absence of reliable redshift 
information for lens galaxies may be difficult. 

We find that the data are rather insensitive to the truncation radius of 
the satellites inside the larger halo as long as it is not very small. 
We can, however, test the extreme possibility that satellites 
have no dark matter attached to them. This is unlikely to be the case
for all satellites even though it may be valid for some fraction 
of those near the center. For example, \cite{2004MNRAS.352L...1G}
argue that up to 40\% of galaxies in clusters may not have associated
DM halos, a conjecture which may be testable with the future weak
lensing measurements (see Fig.~\ref{F:trunc}). 

Additional information about the galaxy-dark matter halo connection may 
be more challenging to extract from the data. 
For example, we have assumed that the number of satellites inside the 
halo scales linearly with halo mass. If this assumption is dropped and 
a more general power law relation is allowed,  
a degeneracy is developed between the satellite fraction and 
the power law slope, but the overall fraction of 
galaxies in $(10^{13}-10^{14})h^{-1}M_{\sun}$ halos is preserved. 

We note here that the power 
law slope and other parameters may be constrained by other observations
such as the galaxy auto-correlation function analysis, velocity 
information in redshift surveys and direct counting of galaxies
in groups and clusters. In fact, a halo model similar to the one 
used here was applied to the  
galaxy auto-correlation analysis \citep{2004astro.ph..8569Z}. 
Combining and comparing the two data sets should provide 
important consistency checks and will
further constrain the galaxy-dark matter connection. 

\section*{Acknowledgements}

RM is supported by the NSF Graduate Research Fellowship Program (NSF
GRFP).  US is supported by a fellowship from the David and Lucile
Packard Foundation, NASA grants NAG5-1993, NASA NAG5-11489 and NSF
grant CAREER-0132953.  AVK and AT are supported by the National
Science Foundation (NSF) under grants No.  AST-0206216 and
AST-0239759, by NASA through grants NAG5-13274 and NAG5-12326, and by
the Kavli Institute for Cosmological Physics at the University of
Chicago.  RHW is supported by NASA through a Hubble Fellowship
awarded by the Space Telescope Science Institute, which is operated by
the Association of Universities for Research in Astronomy, Inc, for
NASA, under contract NAS 5-26555.  The simulations used in this study
were performed on the IBM RS/6000 SP3 system at the National Energy
Research Scientific Computing Center (NERSC).

\bibliography{../BibTeX/apjmnemonic,../BibTeX/cosmo,../BibTeX/cosmo_preprints}
\bibliographystyle{mn2e}

\end{document}